\documentclass{article}[12pt]
\usepackage{epsfig}
\usepackage{supertabular}
\usepackage{appendix}
\usepackage{hyperref}

\begin{document}

\title {Malware ``Ecology'' Viewed as Ecological Succession: Historical Trends and Future Prospects}
\author{Reginald D. Smith \\ PO Box 10051, Rochester, NY 14610 \\ rsmith@citizenscientistsleague.com}
\date{September 24, 2014}

\maketitle

\begin{abstract}
The development and evolution of malware including computer viruses, worms, and  trojan horses, is shown to be closely analogous to the process of community succession long recognized in ecology. In particular, both changes in the overall environment by external disturbances, as well as, feedback effects from  malware competition and antivirus coevolution have driven community succession and the development of different types of malware with varying modes of transmission and adaptability.
Keywords: Ecology, computer virus, community succession, artificial life
\end{abstract}

\newpage
\section{Introduction}
The existence of malware has been almost synonymous with the progress of the personal computer revolution. First, came tentative theories in science fiction works such as \emph{The Shockwave Rider} by John Brunner \cite{shockwave} or \emph{Neuromancer} by William Gibson \cite{neuromancer}. Next were the first proofs of concept by Xerox PARC researchers John Schoch and Jon Hupp. Finally, the first real ``wild'' virus known as Brain emerged in 1986, created by two brothers in Pakistan \cite{virushistory}. Since that time, malware has continued to grow and evolve becoming both a nuisance and a threat. Most recently, it has also become a damaging form of international crime costing \$113 billion in 2013 according to estimates by Symantec \cite{symantec2013} .

Since the very first use of biological analogies such as virus and worm to describe these programs, the analogies and links between computer epidemiology and that of biological epidemics have not gone unnoticed. Indeed, discussions of viruses as artificial life and their similarities to real epidemics are widespread \cite{kephart,pastorvesp, newman, balthrop,serazzi}. Historically, the most detailed research on computer malware has relied either on simulating epidemics to match or revise widely known epidemic models such as the Susceptible-Infected-Recovered (SIR) model or Susceptible-Infected-Susceptible (SIS) model \cite{kephart,pastorvesp} or using theoretical methods and simulation to estimate the epidemic thresholds (or lack thereof) on computer network topologies \cite{pastorvesp,may}.

Many papers have also discussed malware qualitatively in an ecological sense \cite{crandall}. This paper will attempt to add to this discussion by presenting quantitative results and analyses of the evolution of malware over the past 25 years in order to argue that much of malware development can be understood in terms of community succession and that its progressive developments have mirrored succession dynamics in biological ecosystems.

\section{Malware types and their evolution}

To begin our discussion of malware ecology, we must first define malware. Malware is a general term that encompasses a wide variety of malicious programs that can be said to operate on or against computer systems for ulterior motives not approved of or even recognized by users or standard software. Malware encompasses all of the commonly recognized exploits such as computer viruses, worms, trojan horses, and even lesser known malicious programs such as adware and spyware. The main types of malware that will be discussed in this paper, however, are the four most common over the last 25 years shown in Table \ref{malwaretable}.

\begin{table}
\small
\begin{tabular}[b]{|p{0.75in}|p{2in}|p{0.75in}|p{0.75in}|}
\hline
Malware Type&Transmission Method&Approximate Origin Year&Percent of infections in 2013\\
\hline
Computer Virus&Self-replication and passive spread through infection of computer files&1986&9\%\\
\hline
Trojan (Horse)&Spread through user's activities with files or Internet; not self-replicating&1989&87\%\\
\hline
Macro Virus&Self-replication and active spread using Microsoft Office and other software products' macros&1996&0\%\\
\hline
Worm&Self-replication and active spread across media and networks seeking new and unknown hosts&1998&4\%\\
\hline
\end{tabular}
\caption{Major malware types used in the paper, their method of transmission, approximate year of origin and percentage of malware detected by Symantec's Threat Report system in 2013.}
\label{malwaretable}
\end{table}

Granted, these definitions are not always completely mutually exclusive. There can be malware that incorporates aspects of each. For example, a trojan that infects a computer from an email and then proceeds to spread itself on removable media in computer virus-like fashion. However, these four definitions, when considering the main method of transmission, can encompass the vast majority of malware now in existence. In addition to the malware types, malware can be classified into `families' which begin with a single strain but can grow to collectively contain many `variants' which are new versions of the original strain or code sample. Most malware  named in the popular press are families which can have an increasing number of variants as new versions are released to evade security software or exploit new vulnerabilities.

Over time, however, different types of malware have grown, become predominant, and then declined. The actual number of computers infected by different types of malware is extremely difficult to estimate, even with a relatively easily detectable worm such as Conficker \cite{confickerteam}. The estimates of malware infection often depend on either statistics reported from computer security and antivirus products or the monitoring of Internet traffic easily attributable to malicious code over network links. The former method runs into issues of sample size and bias that are difficult to correct while the latter relies on distinguishing malware traffic from all other legitimate traffic with similar features, not always an easy task. Worldwide, about 83\% of computers have antivirus software \cite{mcafee} but it is unknown how many of these actively provide information back to the vendor,additionally, this install base is divided across many different vendors. This leads to other issues such as sample biases where corporate networks or personal computers of certain national or demographic groups are likely overrepresented. Much malware likely lurks with those users not sophisticated enough or without access to actively updated security products.

While overall population estimates are difficult to determine, a proxy for the popularity of types of malware can be found in the creation of new malware of each type, and their derivative variants. Malware writers are an adaptable group and through time the introduction of different types of malware is intimately related to the success of that type of malware in exploiting system weaknesses, transmission, and accomplishing the malware writer's aims. The emergence of new variants by type is relatively easier to measure and while there are still sample size and bias issues, counting new variants only requires proof of existence and not sample size dependent on easily biased statistical measures such as population sizes and statistical moments. Detection by an antivirus system and subsequent reporting as well as ``honeypots'' established by security researchers end up corralling a large portion of all malicious code and definitely most malware that has been able to successfully spread across multiple computers and networks.

The dates of the first appearance of these new variants, while they may vary somewhat in time as reported by different groups, are relatively easy to pinpoint given the short generation time and rapid spread of successful malware. Therefore, the dates of first detection, grouped within a quarterly (three month basis) provide and accurate designation for the age of different malware variants.

\section{Data Sources \& Methodology}

To measure the succession of different types of malware over time, this paper will use data on the emergence of malware variants, grouped by type, over time on a quarterly basis.  This data is available, in various degrees of granularity, from most large antivirus vendors but the data from the Symantec Threat definitions \cite{symthreat} and their annual security reports are used as the primary sources of data for this study.

In the online Threat Definitions, Symantec lists the most common malware families and their major variants by name, type, and often date of first detection. Not all families are listed with data, but for those that were, the data accessed on April 18, 2014 listed a total of 13,599 families and major variants detected over all time. Of these about 44\% were classified primarily as a type of Trojan, 23\% were classified primarily as worms, 13\% as computer or macro viruses, and 20\% as other types of malware. However, on a deeper analysis this list was found to have several features that require revision for a usable dataset.

Notably, were missing detection dates for malware, primarily amongst earlier emerging viruses (before 2000). These were filled in from other sources where data was available (see Appendix). In addition, for clarity, malware counts were only used when they had one, clear definition. Thus a designation such as `Trojan Virus' which indicates a program introduced as a Trojan Horse but subsequently spreading as a virus, was excluded from the overall count of Trojans.

Most impactful, however, was the change in methodology of how threats were tracked over time. Earlier antivirus programs relied on frequently updated lists of exact malware threats to search for and clean. However, with the rapid explosion in malware from 2006 on, where new variants rapidly emerged due to new techniques designed to obfuscate the signature of the malware, new threat definitions emerged which were not tailored for specific named variants but rather were generic algorithms to search for common types of threats using heuristic algorithms. This was acknowledged by Symantec as a reason for the drop in number of new Win32 (targeting 32-bit Microsoft Windows operating system) malware variants detected from 2006 on \cite{symantec2006}. Therefore, these heuristic algorithm threat definitions, such as Bloodhound and Packed.Generic, could not be counted identically to named variants in accounting for the emergence of distinct new malware threats and had to be removed.

One key mistake to not make in enumerating variants of malware is to assume that metamorphic or polymorphic code, malware that can mutate its code as it reproduces, creates new variants. Mutations as they are currently known in malware are used to hide the malware from antivirus detection. While the overall mutation engine of the malware changes the fitness of the malware variant versus others in its family and type, the mutations themselves are more like silent mutations in genetics since they do not confer additional incremental fitness to the malware program and do not create new variants. All new variants so far found have been human designed.

Finally, the data is limited in that not all malware can easily be turned into a named variant for a definition file. In particular, there was an explosion in web exploits from 2006 on, and web threats, which often involve  malicious scripts on websites and other specially tailored code, are not easily assigned as a named threat for a definition file. While general techniques such as cross-scripting can be protected against using more comprehensive ``Internet Security'' solutions rather than just antivirus software, they are not as easily enumerated as other traditional type of threats. As reflected in Figure \ref{malwarethreats}, the emergence of web threats along with the use of heuristic algorithms causes what seems to be a contradictory trend in the data on new threat emergence where there is a sharp, but brief dip in new threat emergence in 2006.

\subsection{Final data set}

With the changes made as described above, a final data set was created with the following attributes: $N$=8,530 different families and major variants classified as one of the four types in Table \ref{malwaretable}. Each has a date of first detection from either Symantec or another named source (see appendix). The dates range from the Brain virus in 1986 to the latest threats of the fourth quarter of 2013. Data from 2014 was removed since trend data indicated it was largely incomplete and showed a marked decrease in overall number of threats from earlier periods, likely due to a lag in reporting new threats publicly.

\section{Analysis of Malware Over Time}

\begin{figure}
\centering
 \begin{tabular}{cc}
 	 \includegraphics[height=2.5in, width=2.5in]{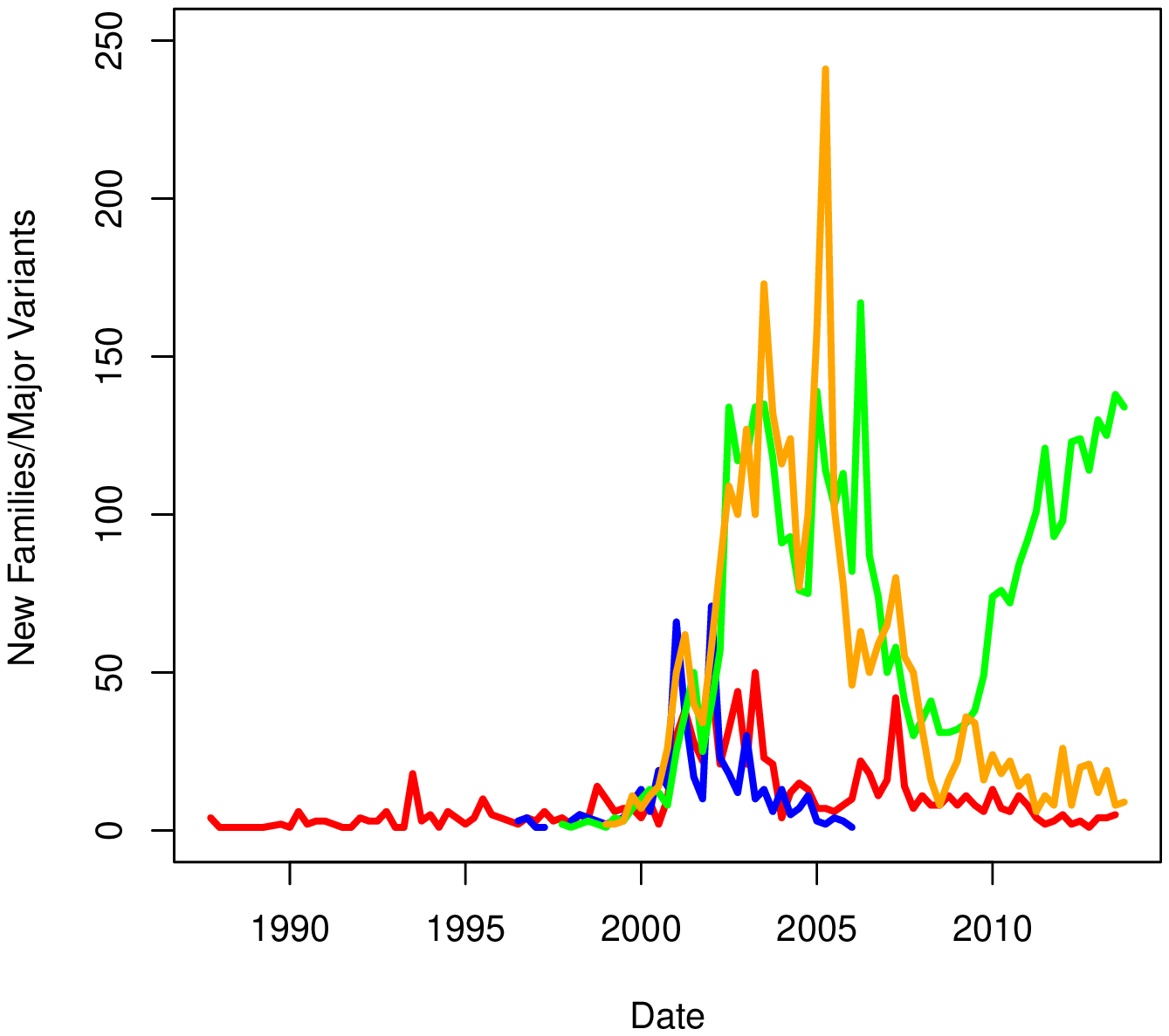}&
    \includegraphics[height=2.5in, width=2.5in]{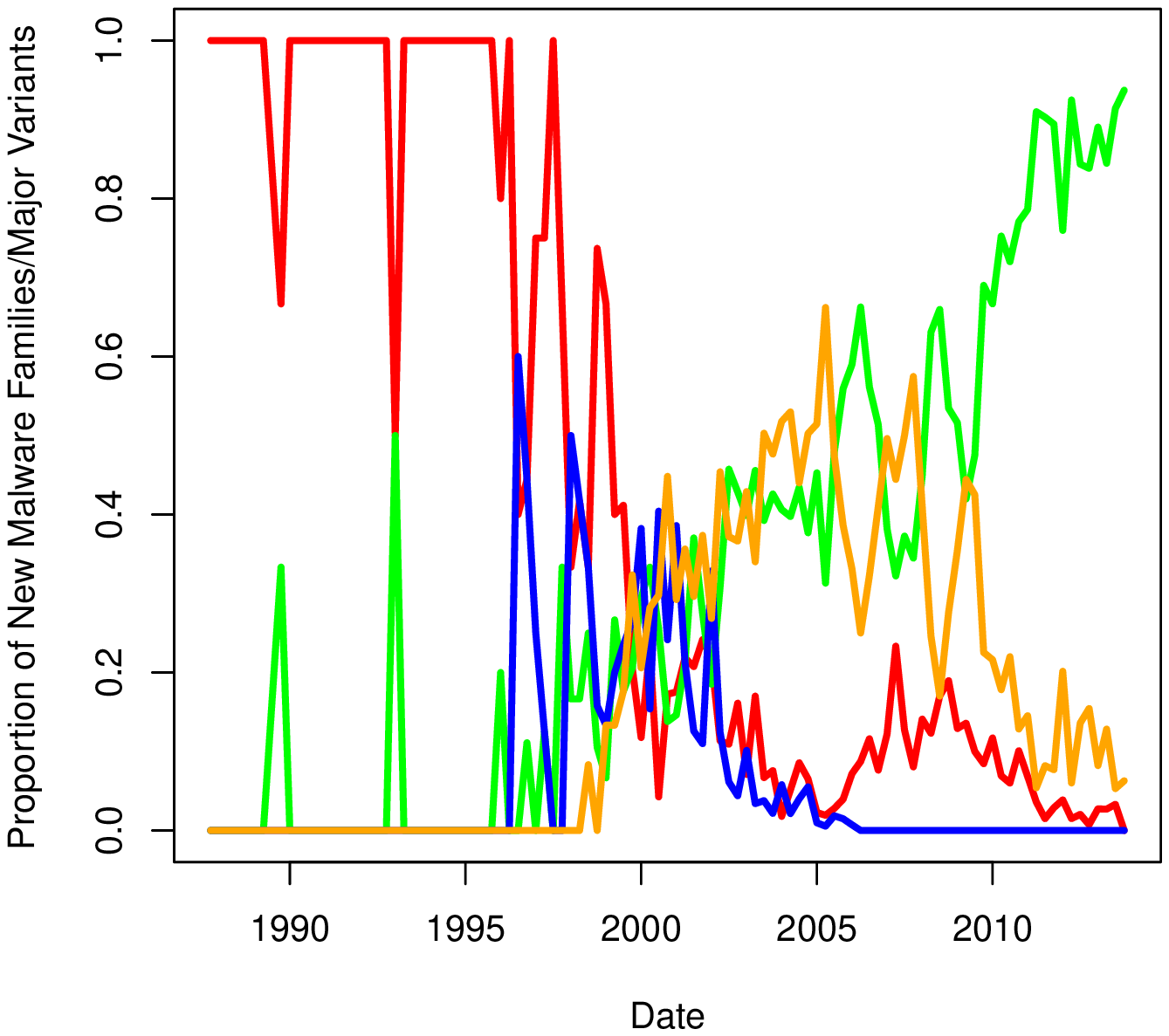}\\
 \small{New malware threats by type over time}&\small{Share of new malware threats by type over time}\\
\end{tabular}
\caption{The number of new malware threats by type over time from 1986 to 2013 and the share of new malware threats by each type over the same period. Colors for each malware type are red for computer viruses, green for Trojans, blue for macro viruses, and orange for worms.}
\label{malwarethreats}
\end{figure}

In Figure \ref{malwarethreats}, the evolution of malware threats over time is shown both as a graph of the total number of new threats per quarter as well as the proportion of all new threats represented by each type of malware per quarter. As is initially strikingly clear, malware has gone through multiple stages where different types of malware have been most prominent at different times.

Both metrics are valuable, though proportion is likely more so. Because of the unknown bias in sampling, coverage of Symantec's threat data, etc. the actual count, while probably accurate, cannot be considered complete. Further, the families and variants named are only those which Symantec thought were important enough to name and give a date. Proportion on the other hand allows us to measure the relative strength and popularity of each malware type relative to the others across time. As long as there are no huge gaps that bias detection of one type of malware over another, these proportions are probably relatively accurate.

Using the proportion data, it is clearly seen that there are several key stages in malware evolution. In the first, from 1986 to about 1996, computer viruses, particularly the disk boot sector virus, ran the show. There was the occasional Trojan as well but the default method of transmission, via infected floppy disk, did not change much over time. Starting in 1996, with the introduction of macros, short programs for routine functions, in Microsoft Office, a new form of virus emerged: the macro virus. By executing malicious code within Microsoft Office it could cause havoc, as well as spread itself through channels such as disks or emails via the infected computer's Outlook address book. The macro virus burst on the stage and rapidly grew to become the fastest increasing threat of that time and briefly the largest source of new infections.

The years 1998 to 2002 stand out as an era of great diversity amongst malware and a lack of dominance by any type. The computer worm, which actively propagated itself without user intervention or the need of a specialized program like MS Office, debuted in 1998 and spread rapidly with several large, global infections such as Code Red, Slammer, and Nimda. Viruses continued to play a major role though one of their main vectors of propagation was in rapid decline with the coming obsolescence of the floppy disk.

\begin{figure}
\centering 
\includegraphics[width=2.5in, height=2.5in]{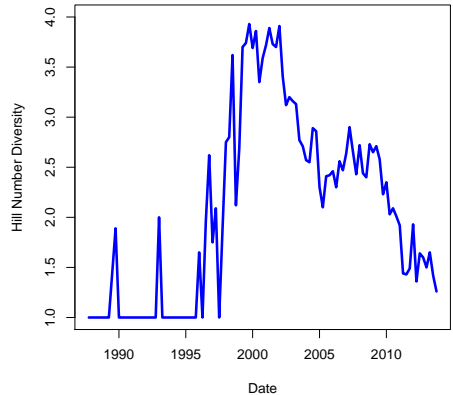}
\caption{Graph of the diversity of new malware across the four main types over time from 1986 until the end of 2013. Diversity is calculated using the Hill numbers where $a$=1.}
\label{hillchart}
\end{figure}

The quantitative measure of malware diversity during this time can be seen in Figure \ref{hillchart} where the diversity is measured as the Hill Number based on the proportion of new malware by each type every quarter. Hill Numbers have been shown to be one of the best and most consistent indicators for diversity \cite{hill1,hill2,hill3}. The basic Hill Numbers of order $a$ are defined as

\begin{equation}
N_a = (p_1^a+p_2^a\dots+p_n^a)^{1/(1-a)}
\label{hillgeneral}
\end{equation}

In equation \ref{hillgeneral}, $p$ is the proportion of each species by number of individuals (or dry mass for plants) to the entire community population while $a$ represents the order of the Hill Number. Different orders represent different aspects of the population where $a=0$ is the number of species, $a=1$  the diversity, and the inverse of $a=2$ is Simpson's number.\  Hill showed in \cite{hill1} that the Hill numbers are in effect the R\'{e}nyi entropy of order $a$. In this paper the Hill number for $a=1$ is used and is calculated as

\begin{equation}
N_1 = \exp\bigg(\sum_{i=1}^{4} -p_i \log{p_i}\bigg)
\end{equation}

As can be seen in Figure \ref{hillchart}, the diversity of the malware ecosystem surged from 1998 forward peaking between 2000 and 2002. The proportion of each type of malware is roughly equal over this time period. From 2002 on, however, the diversity begins a slow decline. The first reason is the rapid extinction of the macro virus. Security features in the Microsoft Office products such as prompts before the automatic execution of macros and improvements in antivirus software to detect macro virus threats made macro viruses much less dangerous and more difficult to propagate. Ten years after their debut, in 2006, they disappeared entirely as a major source of malware.

The second reason, alluded to previously, was the decline of floppy disks. This eliminated one of the main vectors for computer viruses and they declined accordingly. They would recover, with the ability to exploit media such as USB keys or mobile devices, but would never again be the dominant force they had been in the early days of personal computing. 

The field was largely left to both worms and trojans. Worms, however, began their own long-term decline starting in 2007-2008. Many of the major worm epidemics had relied on major security vulnerabilities in operating systems or software to infect other computers and propagate. The security flaw in the Microsoft 2000 Server IIS (Internet Information Server) was a key factor in the incredibly rapid spread of Code Red. In addition, the nature of the Internet topology, with a power law or scale-free distribution of connections, made it so that there was theoretically almost no epidemic threshold \cite{pastorvesp, may}. With the plugging of these vulnerabilities as well as improved antivirus detection and removal programs, massive worm epidemics became a thing of the past. One of the last major epidemics, and one of the worst, were the Conficker worms which peaked in 2008.

From this point, the malware diversity continued its slow decline until reaching levels last seen in 1997. Trojans are by far the major source of new malware, partially because of their flexibility in relying on targeting specific user behaviors and programs rather than system wide attacks like viruses or worms. In addition, their range of forms may make it so that the Trojan category is overused to describe all types of malware that do not propagate between users.  Trojans have also become dominant since they are a key method of committing cybercrime such as creating botnets, stealing account or password information, or installing 'ransomware' that causes computer malfunction until a ransom is paid as directed by the software.

\subsection{Malware ecology and community succession}

Community succession is one of the most widely known and studied processes in ecology. It is defined as the temporal change in species composition over time in a fixed geographic area. In succession, the mix of species in a similar geography evolves over time due to both external disturbances such as natural disasters or man-made destruction as well as due to changes in the environment by prior species. In community succession, the ecology of the community evolves in several general stages.

\begin{enumerate}
\item  Primary Succession - Where a group of species colonizes a previously uninhabited (i.e. new) geographic area

\item Secondary Succession - The arrival and colonization of new groups of species after a disturbance has largely removed the primary colonization species. These new species can take advantage of the environment as it has been changed by the primary species

\item Additional successions can also occur given environmental variables or invading species.

\end{enumerate}

A classic example of community succession was given by the succession of plant life on and around the Bikini and Enewetak Atolls after the thermonuclear weapons tests beginning in 1954 \cite{bikini1,bikini2,bikini3,bikini4}. In \cite{bikini1, bikini2}, Held and Palumbo describe how the initial blast eliminated almost all vegetation but that shrubs had reappeared within two years time, mainly from seeds or old stumps of \emph{Scaevola frutescens} and \emph{Messerschmidia argentea}. In \cite{bikini3}, the trees \emph{Pisonia Grandis} are mentioned as often replacing coconut palms, particularly after the devastation of a typhoon or nuclear test.

Malware has followed a similar pattern from the origin of the first virgin, ecological environment, the early personal computer, and its ability to share data through floppy boot disks. The rise of viruses on personal computers also led to the rise and coevolution of antivirus software. Computer viruses responded in ways that pointed to increasing sophistication such as polymorphic and metamorphic code, which mutated code to block antivirus software looking for signatures, and encryption. With the rise of new technologies such as macros and the Internet, additional avenues for propagation became available. 

This secondary succession of macro viruses, worms, and trojans was based on the previous success of the development of advanced computer virus code in addition to new technology. However, the spectacular success of some macro viruses and worms led to their destruction since the software and antivirus vendors were able to respond to and close these powerful, though very specific, software vulnerabilities leading to their decline. Macro viruses, in particular were destroyed by the removal of their specific niche in productivity software macros when automatic script execution was prevented and other safeguards put into place. Similarly, the progress of technology that allowed secondary succession obsoleted the floppy boot disk and eliminated a main niche of the boot sector virus.

\section{Conclusion}

The diversity amongst malware has gone through several stages which mirror an ecological succession in ecosystems. From starting with only boot sector computer viruses in the 1980s, the numbers and diversities rapidly rose through the early 2000 era as macro viruses, worms, and trojans were introduced. Changes in the environment, often brought on by the very success of advanced malware, began to eliminate niches for those such as macro viruses or epidemic worms while changes in data transfer practices with the end of floppies severely restricted traditional computer viruses. While the current environment seems dominated by Trojans, this is largely due to them having one of the more user targeted and heterogeneous approaches of all malware. This makes large-scale patches and other mass fixes less effective and allows Trojans to perpetuate for a long time. In fact, the decline in malware type diversity likely masks a broader intra-Trojan diversity.

This does not necessarily herald the end of malware innovation. Several trends are likely to manifest in the future altering the landscape yet again:

\begin{enumerate}
\item The increasing use of mobile devices, tablets, home and car automation, and the broader Internet of things will provide a fertile environment for malware with multiple methods of propagation and flexibility across devices. This has not yet extensively been seen in regular malware but is one of the features of recent suspected state-developed malware such as Stuxnet.
\item Continued blurring of lines between malware types as propagation methods and key techniques are shared and integrated into new types of malware.
\item The ability of malware to undergo non-user directed evolution. Whereas polymorphic and metamorphic code currently only increases malware fitness by the skill of the mutation engine, a dangerous development in future malware may be the ability undergo mutations similar to biological processes where the mutation alters the function of the malware itself and thus changes its fitness with each mutation. 
\item The former point would accelerate the development of advanced antivirus software beyond heuristics to a more adaptive detection system similar to immune response in biological systems.
\end{enumerate}
Therefore, the current low diversity and development of new types are likely a brief interlude and not the end of the story.

\newpage
\begin{appendices}

\begin{table}
\section{Appendix - Other Sources of Dates}
\small
\begin{tabular}[b]{|p{2in}|p{2.5in}|}
\hline
Source&Location\\
\hline
Patricia Hoffman's Virus Information Summary List (VSUM)&\url{wiw.org/~meta/vsum}\\
\hline
WildList Organization International&\url{wildlist.org}\\
\hline
Antivirus DownloadAtoZ Virus Definitions&\url{antivirus.downloadatoz.com}\\
\hline
Wikipedia&\url{www.wikipedia.org}\\
\hline
McAfee&\url{www.mcafee.com}\\
\hline
H.J. Highland ``A History of Computer Viruses''&Highland, H.J. (1997). ``Special feature: A history of computer viruses - Introduction.'' Computers and Security, 16:5, 412-415\\
\hline
Virus Information (wikia)&\url{virus.wikia.com}\\
\hline
Fauzi, ``A Study on Computer Viruses\dots ''&Fauzi, Mohd, Sanim, Shukor (2003) A study on computer viruses attacks and the way to cure them / Shukor Sanim Mohd Fauzi. Bachelor Degree thesis, Universiti Teknologi MARA (UiTM). \\
\hline
Virus Information Update CIAC (Computer Incident Advisory 
Capability)&\url{ciac.llnl.gov}\\
\hline
The Security Digest&\url{SecurityDigest.org}\\
\hline
Risks Digest&\url{www.risks.org}\\
\hline
Lower Columbia College Y2K Viruses List&\url{www.lcc.ctc.edu/info/y2k/y2kvirus.xtm}\\
\hline
CA Associates Anti-Spyware&\url{www.pestpatrol.com}\\
\hline
\end{tabular}
\end{table}
\end{appendices}

\end{document}